\documentclass[fleqn,twoside]{article}
\usepackage{espcrc2, epsfig}
\title{Relativistic bound-state calculations in Light Front Dynamics}
\author{M. van Iersel, B.L.G. Bakker, and F. Pijlman \\
	Department of Physics and Astronomy \\
	Vrije Universiteit, Amsterdam, the Netherlands}
\begin{document}
\begin{abstract}
We calculated bound states in the quantum field theoretical approach.
Using the Wick-Cutkosky model and an extended version of this model
(in which a particle with finite mass is exchanged) we have calculated
the bound states in the scalar case.
\end{abstract}
\maketitle
\section{Introduction}
There are several ways of calculating the bound states of a system in the
context of quantum field theory. We have chosen to calculate the bound states
in a quantum field theoretical approach in Light Front Dynamics (LFD), since
LFD is very well suited for this problem (see e.g. \cite{bak}, \cite{brodpaul}).

Our goal is to move towards QCD, but before treating a realistic system we
have done calculations in the scalar case. In the case that a massless particle
is exchanged ($\mu = 0$) we treat the Wick-Cutkosky model. An extended version
of this model is used in the case that a particle with mass ($\mu \neq 0$) is
exchanged. Work on this subject has already been done by others; e.g.
\cite{glaz,tritt,mangin}.

After setting up the equation and explaining which method we use to solve it
(section 2), we will present some results in section 3. At the end we will give a
short conclusion and outlook.
\section{Setting up the problem and solving it}
The physical system we have looked at consists of two scalar particles of
equal mass $m$ exchanging another scalar particle. This particle can either
have a mass equal to zero ($\mu = 0$) or it can have a mass $\mu$ unequal
to zero. The Lagrangian for this system is given by:
\begin{eqnarray}
\mathcal{L} &=& \frac{1}{2} \partial_{\mu}\Psi \partial^{\mu}\Psi - \frac{m^2}{2}
\Psi^{2} + \frac{1}{2} \partial_{\mu} \Phi \partial^{\mu} \Phi \nonumber\\
&& - \frac{\mu^2}{2} \Phi^{2} - g\Phi\Psi^{2} \, ,
\label{eqn1}
\end{eqnarray}
where $g$ is the coupling constant.

\begin{figure}[h]
\vspace{-5mm}
\epsfig{figure=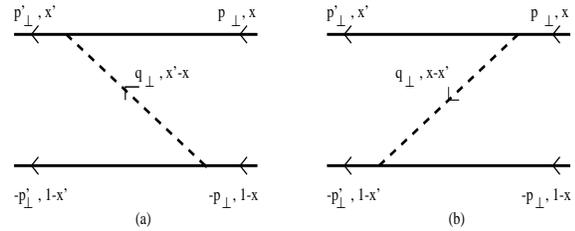,width=7.5cm, height=3cm, angle=0}
\caption{First order time-ordered diagrams in the ladder approximation
for one particle exchange.}
\label{todiagram}
\vspace{-5mm}
\end{figure}

Taking along only two and three particle states, we can derive the
bound-state equation from this lagrangian. It is given by:
\begin{eqnarray}
\left[ M^{2} \!\!\!\!\right.&-&\left.\!\!\!\! \frac{\vec{p}_{\perp}^{\,2}
+m^{2}}{x(1-x)} \right] \Psi(\vec{p}_{\perp},x) = \nonumber\\
&& \hspace{-1cm} \frac{g^{2}}{(2\pi)^{3}} \int\!\! \mathrm{d}^{2}
\vec{p}_{\perp}^{\,\prime}\mathrm{d}x^{\prime}
K(\vec{p}_{\perp},x;\vec{p}_{\perp}^{\,\prime},x^{\prime})
\Psi(\vec{p}_{\perp}^{\,\prime},x^{\prime}) \, .
\label{eqn2}
\end{eqnarray}
Here $M$ is the total mass of the system and $K(\vec{p}_{\perp},x;
\vec{p}_{\perp}^{\,\prime},x^{\prime})$ is the kernel of the equation.
There are several possible expressions for the kernel, depending on
what approximation we make. Here we have been working in the ladder
approximation. In first order we have two time ordered diagrams
(Fig.~\ref{todiagram}).
Each of these diagrams corresponds to an energy denominator which is
present in the kernel of the bound-state equation (Eq.~(\ref{eqn2})).
The total kernel contains, besides the energy denominators, a
phase-space factor and a theta function, which determines in which
time-ordering we are working.\\

\noindent For the total kernel we can write:
\begin{eqnarray}
K(\vec{p}_{\perp},x;\hspace{-1.3cm}&\hspace{-1.2cm}\vec{p}_{\perp}^{\prime},x^{\prime})
= \nonumber\\
& \hspace{-2mm} \frac{1}{\Phi(x,x')} \left( \frac{\theta(x'-x)}{2(x'-x)D_{a}} +
\frac{\theta(x'-x)}{2(x-x')D_{b}} \right) \, .
\label{eqn3}
\end{eqnarray}
Here $\Phi(x,x')$ is representing the phase-space factor, which is given by:
\begin{eqnarray}
\Phi(x,x') = \sqrt{x(1-x)x'(1-x')} \, ,
\label{eqn4}
\end{eqnarray}
and $D_{a}$ and $D_{b}$ are the energy denominators corresponding to left or
right graph in Fig.~\ref{todiagram}. The expressions for these energy
denominators are:
\begin{eqnarray}
D_{a} &=& M^{2} - \frac{\vec{p}_{\perp}^{\,2}+m^{2}}{x} -
\frac{\vec{p}_{\perp}^{\,\prime\,2} + m^{2}}{1-x'} \nonumber\\
&&- \frac{(\vec{p}_{\perp}^{\,\prime} - \vec{p}_{\perp})^{2} + \mu^{2}}{x'-x}
\, , \label{eqn5} \\
D_{b} &=& M^{2} - \frac{\vec{p}_{\perp}^{\,\prime\,2}+m^{2}}{x'} -
\frac{\vec{p}_{\perp}^{\,2} + m^{2}}{1-x} \nonumber\\
&&- \frac{(\vec{p}_{\perp} - \vec{p}_{\perp}^{\,\prime})^{2} + \mu^{2}}{x-x'}\, .
\label{eqn6}
\end{eqnarray}
The bound-state equation has been solved by making an expansion into
basis functions, symmetrizing it and then using Gaussian integration
to solve it. The wave functions depend on both $x$ and $|\vec{p}|$
and we have used two different functions to expand in. For the $x$ part
of the wave function we have used cubic spline functions and for the
$|\vec{p}|$ part we have used a basis which was first introduced by
Olsson \cite{ols} and Weniger \cite{wen}. We have tested the latter
basis in instant form \cite{ier} and we have seen that it works well.

\section{Results}
Using the method described in the previous section, we are able to calculate
the spectra and wave functions in case of massless exchange as well as in the
case of massive exchange. Here we have to remark that we did not take into account
any higher order terms in the Hamiltonian or self energy diagrams.

First of all we have looked at how many functions we need to take along to reach
a reasonable convergence. A reasonable number of spline functions to take into
account in the calculations is 10. Although 8 spline functions is a bit on the low
side, it already gives good results as well. For the Olsson Weniger functions
we need to take 3 functions to reach a good convergence. Taking less functions
already gives a good convergence in the lower states, but not in the higher excited
states. Note that increasing the number of Olsson Weniger functions in the expansion
causes a large increase in computer time. Due to this we have chosen to use 8
spline functions and 3 Olsson Weniger functions in the expansion to obtain the
results presented here.

\begin{figure}
\vspace{-2cm}
\epsfig{figure=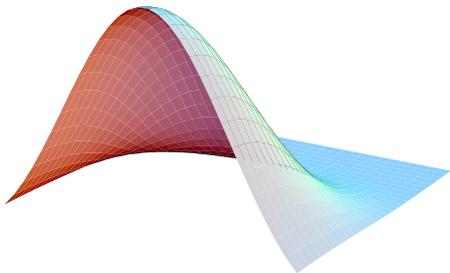,width=6cm, height=5cm, angle=0}
\caption{Ground state wave function for massless exchange.}
\label{ground1}
\end{figure}

\begin{figure}
\vspace{-2cm}
\epsfig{figure=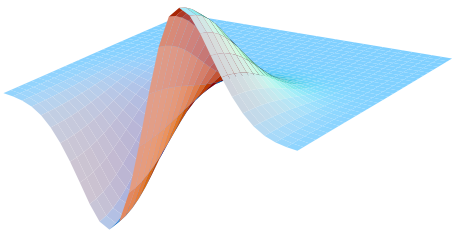,width=6cm, height=5cm, angle=0}
\caption{Wave function of the first excited state in the case of
massless exchange.}
\label{excited11}
\end{figure}

\begin{figure}
\vspace{-2cm}
\epsfig{figure=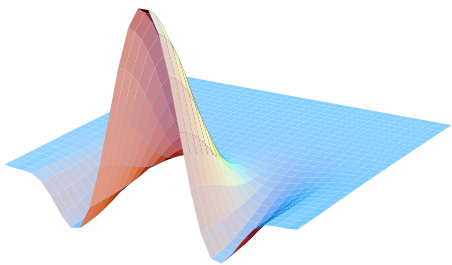,width=6cm, height=5cm, angle=0}
\caption{Wave function of the second excited state in the case of
massless exchange.}
\label{excited12}
\end{figure}

\begin{figure}
\vspace{-2cm}
\epsfig{figure=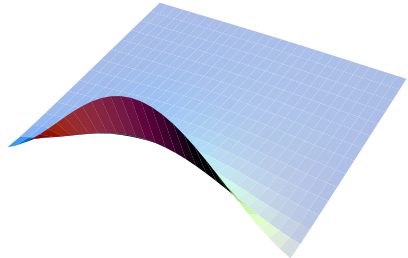,width=6cm, height=5cm, angle=0}
\caption{Ground state wave function in the case of massive exchange ($\mu = 0.15$).}
\label{ground2}
\end{figure}

In the case of massless exchange ($\mu = 0.0$) we have calculated the masses given
in the second column of table~\ref{mass}. In the third column of this table the
masses of the bound states are given in the case of massive exchange ($\mu = 0.15$).
In both cases we have used a mass $m = 1.0$ for the particles and a coupling
constant of $g = 17.36$.
\vspace{-5mm}
\begin{table}[h]
\begin{center}
%\caption{Masses corresponding to the bound states calculated in the case of massless
%and massive exchange using the following parameters: $m = 1.0$ and $g = 17.36$.}
\begin{tabular}{|c|c|c|}
\hline
n & mass ($\mu = 0.0$) & mass ($\mu = 0.15$) \\
\hline
0 & 0.75 & 0.91 \\
1 & 1.63 & 1.75 \\
2 & 1.74 & 1.86 \\
3 & 1.86 & - \\
4 & 1.91 & - \\
\hline
\end{tabular}
\vspace{3mm}
\caption{Masses corresponding to the bound states calculated in the case of massless
and massive exchange using the following parameters: $m = 1.0$ and $g = 17.36$.}
\label{mass}
\end{center}
\vspace{-12mm}
\end{table}
We have also calculated the wave functions corresponding to the masses given
in table~\ref{mass}. For the massless case the wave functions corresponding
to the first three states are given in Fig.~\ref{ground1}, \ref{excited11}
and \ref{excited12}.
The wave function corresponding to the ground state in the case of massive exchange
($\mu = 0.15$) is given in Fig.~\ref{ground2}.
Looking at these figures one should notice that the wave functions are not
drawn in the ordinary way. This is due to the fact that the wave function
depends on $x$ (the fraction of the $p^{+}$-momentum) and $\vec{p}_{\perp}$,
instead of $\vec{p}$ as is usually the case.
\section{Conclusions and outlook}
The bound-state equation in the scalar case is solved by making an expansion
into basis functions. The basis we have used is well suited for this problem
and we expect it to work in more realistic cases as well. With this method we
can calculate the ground state as well as excited states in both cases, massless
and massive exchange.

The plans for the future are to move towards a more realistic system, i.e. to
include spin. And also to included higher order terms in the Hamiltonian and see
whether these terms give a large contribution.

\begin{thebibliography}{50}
\bibitem{bak} B.L.G. Bakker, Lecture notes in Physics, Vol. 572, Springer-Verlag
\bibitem{brodpaul} S.J. Brodsky, H.C. Pauli and S.S. Pinsky,
Phys. Rep. \textbf{301}, 299-486 (1998)
\bibitem{glaz} S. G{\l}azek, A. Harindranath, S. Pinsky, J. Shigemitsu,
and K. Wilson, Phys. Rev. \textbf{D47} 1599-1619 (1993)
\bibitem{tritt} U. Trittmann, and H.C. Pauli, hep-th/9705021
\bibitem{mangin} M. Mangin-Brinet, and J. Carbonell, Phys. Lett.
\textbf{B474} 237-244 (2000)
\bibitem{ols} M.G. Olsson, S. Veseli, and K. Williams, Phys. Rev.
\textbf{D52}, 5141 (1995)
\bibitem{wen} E.J. Weniger, J. Math. Phys. \textbf{26}, 276 (1985)
\bibitem{ier} M. van Iersel, C.F.M. van der Burgh, and B.L.G. Bakker,
hep-ph/0010243
\end{thebibliography}
\end{document}